\documentclass[tighten,10pt,A4,twocolumn,prl]{revtex4}%
\usepackage{amssymb}
\usepackage{graphicx}
\usepackage{amsmath}
\usepackage{amsfonts}%
\providecommand{\U}[1]{\protect\rule{.1in}{.1in}}
\providecommand{\U}[1]{\protect \rule{.1in}{.1in}}
\providecommand{\U}[1]{\protect \rule{.1in}{.1in}}
\providecommand{\U}[1]{\protect \rule{.1in}{.1in}}
\providecommand{\U}[1]{\protect \rule{.1in}{.1in}}
\providecommand{\U}[1]{\protect \rule{.1in}{.1in}}
\providecommand{\U}[1]{\protect \rule{.1in}{.1in}}
\providecommand{\U}[1]{\protect \rule{.1in}{.1in}}

\begin{document}
\title[ ]{Anisotropy in magnetoelastic effects of YCo$_{5}$ }
\author{Jun-ichiro Inoue$^{1}$ and Hiroki Tsuchiura$^{2}$}
\affiliation{$^{1}$Faculty of Engineering, Nagoya University, Nagoya 464-8603, Japan}
\affiliation{$^{2}$Department of Applied Physics, Tohoku University, Sendai 980-8579, Japan}

\begin{abstract}
We develop the phenomenological theory for uniaxial ferromagnets, and derive
explicit formulae for their magnetostriction (MS) and magnetic anisotropy
energy (MAE) generated by the magnetoelastic (ME) effects. An analysis of the
experimental results for the MS in hexagonal YCo$_{5}$ using the formulae
obtained reveals that 1) an elastic anomaly should appear in the elastic
constant $c_{33}$, and 2) the MAE caused by the ME effects could be
non-negligible. The calculated results of electronic structure under lattice
deformation indicate that the particular arrangement of Co atoms in the
lattice is related to the anisotropy of MS in YCo$_{5}$. It is suggested that
the anisotropic MS observed for various Y-compounds could be interpreted in a
similar manner.

\end{abstract}
\maketitle
\date{}

Ferromagnetic intermetallic compounds constituted of Y or rare-earth (RE)
elements with transition metal elements Fe or Co are basic materials for
permanent magnets with large uniaxial magnetic anisotropy (MA)\textbf{
}\cite{kirchmayr79,buschow80,sagawa84,inoue88,herbst91}\textbf{. }Among
them,\textbf{ }YCo$_{5}$ is a strong ferromagnet with a Curie temperature of
approximately 1000 K and saturation moment of 8.2 $\mu_{B}$ per formula unit
(f.u.). The compound is a unique ferromagnet in that several magnetic
properties are highly anisotropic: large magnetic anisotropy energy (MAE)
$K_{u}=3.8$ meV per f.u. \cite{buschow73,sankar75,alameda83}, anisotropic
variation in the lattice constants under a high pressure
\cite{rosner06,koudela08}, and anisotropic magnetostriction (MS)
\cite{andreev95}.

Koudela et al. \cite{rosner06,koudela08} performed a high pressure study on
YCo$_{5}$. They measured the variations in the volume and $c/a$ ratio. Here,
$a$ and $c$ are the lattice constants. They observed that a metamagnetic
transition (MMT), a first order transition between magnetic states with high
and low magnetic moments, occurs at a pressure of approximately 18 GPs. The
observed data reveal that the variation in the lattice constants in the MMT is
anisotropic: $c$ varies more rapidly around 18 GPs than $a$.

Andreev's group reported experimental values of the spontaneous MS for many
intermetallic compounds by measuring the temperature dependence of the lattice
constants for single crystals \cite{andreev95}. A few of the MS constants
observed are presented in TABLE I. Here, $\lambda_{a}$ and $\lambda_{c}$ are
the variations in the lattice constants along the $a$ and $c$ directions,
respectively. As presented in TABLE I, $\lambda_{a}\ll\lambda_{c}$ for
YCo$_{5}$, that is, the MS is significantly anisotropic.

\begin{table}[tb]
\caption{Observed magnetostriction (MS) constants, $\lambda_{a}$ and
$\lambda_{c}$, along the $a-$ and $c-$axes, respectively, in certain hexagonal
(h-), rhombohedral (rh-), and tetragonal (tr-) Y-compounds \cite{andreev95}. }
\begin{center}%
\begin{tabular}
[c]{lll}\hline\hline
material & $\ \ \lambda_{a}$ \ \  & $\ \ \ \lambda_{c}$ \ \ $\ (\times
10^{-3})$\\\hline
h-YCo$_{5}$ & \ \ 0 & \ \ 6.8 \ \ \ \\
h-Y$_{2}$Fe$_{17}$ & \ \ 1.50 & \ \ 8.04\\
rh-Y$_{2}$Fe$_{17}$C$_{1.5}$ & \ \ 6.00 & \ \ 8.26\\
tr-Y$_{2}$Fe$_{14}$B & \ \ 8.55 & \ \ 2.77\\\hline\hline
\end{tabular}
\end{center}
\end{table}

The large MAE of YCo$_{5}$ has been explained in first principles by
considering the orbital polarization in the formalism to recover the Hund's
rule coupling
\cite{coehoorn91,nordstrom92,daalderop96,yamaguchi97,steinbeck01}. It was
indicated that the particular arrangement of Co atoms in the lattice might be
crucial to the large MAE \cite{larson03a,larson03b}.

\begin{figure}[tb]
\begin{center}
\includegraphics[width=0.65\linewidth]{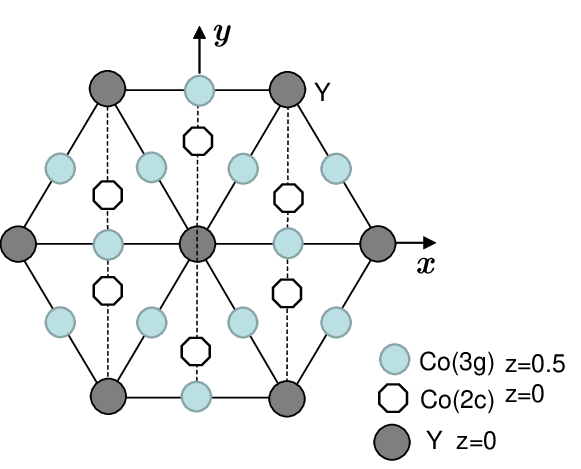}
\end{center}
\caption{(color online) Lattice structure of hexagonal YCo$_{5}$ projected
onto the $a-b$ plane. Co(2c) and Co(3g) sites form honeycomb and kagome
lattices on the $z=0$ and $z=0.5$ planes, respectively.}%
\label{figure1}%
\end{figure}

The occurrence of the MMT in YCo$_{5}$ was predicted earlier in the linear
muffin in orbital (LMTO) calculation by Yamada et al. \cite{yamada99}. Koudela
et al. \cite{rosner06,koudela08} calculated the electronic structure using
first principles, and clarified that a sharp peak of the up-spin density of
states (DOS) causes the MMT.

Hexagonal YCo$_{5}$ includes two non-equivalent Co sites, the 2c and 3g sites
shown in Fig. \ref{figure1}. The 2c and 3g sites form honeycomb and kagome
lattices, respectively, on the $a-b$ plane. The alternate stacking of these
lattices along the $c-$axis generates a pyrochlore-like lattice. It has been
indicated that the particular arrangement of Co sites in YCo$_{5}$ produces
two flat bands \cite{koudela08,ochi15}. One is formed by the $xy$ and
$x^{2}-y^{2}$ orbitals on the Co(3g) sites on the kagome lattice, and the
other is formed by the $xz$ and $yz$ orbitals on the pyrochlore-like lattice.
Because the former flat band gives an invisibly small DOS peak at the top of
the Co d-bands, the sharp DOS peak, responsible of the MMT, is caused by the
flat energy band composed of the\textbf{ }$xz$\textbf{ }and\textbf{ }%
$yz$\textbf{ }orbitals.

The theoretical analysis of the MMT in YCo$_{5}$ shown above indicates that
the characteristic lattice structure of YCo$_{5}$ is responsible for the
anisotropic change in the electronic structure caused by lattice deformation.
Thus, we generally expect that the magnetoelastic (ME) effects of
YCo$_{\mathbf{5}}$ can also be anisotropic. For example, the relations between
the magnetization direction and lattice deformation, known as MS in general,
could be anisotropic. The dependence of the magnetization direction on the
lattice deformation may also induce an additional MAE, called MAE induced by
ME effects. So far, however, no theoretical analysis of the anisotropy in ME
effects in YCo$_{5}$ has been performed.

In the present work, we focus our attention on the anisotropic MS in YCo$_{5}$
and clarify a likely origin of the anisotropy. To achieve this, we develop a
phenomenological theory for uniaxial ferromagnets expressed in terms of the
elastic and ME constants by extending our previous studies
\cite{inoue13,inoue19}. Only the phenomenological theory can treat the ME
effects properly at present; however, the various physical constants such as
the elastic and ME constants may in principle be estimated through electronic
structure calculations without lattice deformation.

We combine the phenomenological analysis and the experimental result
$\lambda_{a}\sim0$ for YCo$_{5}$ presented in Table I, and find that a simple
relation $\lambda_{c}=B_{3}/c_{33}$ holds for YCo$_{5}$, where $B_{3}$ and
$c_{33}$ are respectively the ME and elastic constants along the
$c-$axis.\textbf{ }To estimate the value of $B_{3}$, we calculated a change in
MAE caused by lattice deformation along the $c-$axis by performing electronic
structure calculations, including spin--orbit coupling (SOC)
\cite{inoue15a,inoue21}. Combining the obtained results, we demonstrate that
the elastic anomaly predicted to occur in YCo$_{5}$ \cite{koudela08} should be
caused by softening of the lattice along the $c-$axis, that is, a decrease in
the elastic constant $c_{33}$. We also show that the MAE induced by ME effects
could be non-negligible.\textbf{ }

In the following, we develop the phenomenological theory in Section II and
apply the results to the experimental results of YCo$_{5}$. We present the
calculated results of the deformation dependence of the DOS and MAE of
YCo$_{5}$ in Section III, and compare the results with those obtained in first
principles. In Section IV, we perform a simple analysis on the magnitude of
$B_{3}$, estimate the contribution to MAE by the ME effects, and discuss the
results. The implications of the present results will be presented at the end
of the discussion. Section V summarizes the work. Some details of the
phenomenological theory and results obtained from first principles are
presented in the Appendix.

\section{II. Phenomenological theory}

In this section, we present explicit expressions of MS and MAE obtained using
the phenomenological theory for uniaxial ferromagnets with hexagonal and
tetragonal lattices. A few details are provided in the Appendix.

In the phenomenological theory, the free energy $F$ is given by the sum of the
elastic energy $F_{el}$ and ME energy $F_{me}$. These are expressed in terms
of the elastic constants, lattice deformation, and ME constants. When we use
the elastic tensor $\boldsymbol{C}$, vector representations of the deformation
tensor $\boldsymbol{e}$, and ME constants $\boldsymbol{B}$, the free energy is
given as
\begin{equation}
F=F_{el}+F_{me}%
\end{equation}
with
\begin{align}
F_{el}  &  =\boldsymbol{e}^{t}\cdot\boldsymbol{C}\cdot\boldsymbol{e}%
/2,\label{Fel}\\
F_{me}  &  =\boldsymbol{e}^{t}\cdot\boldsymbol{B}, \label{Fme}%
\end{align}
where the components of $\boldsymbol{B}$, $\boldsymbol{C}$, and
$\boldsymbol{e}$ are given in Appendix A. The expression for $\boldsymbol{B}$
includes the magnetization direction $\boldsymbol{\alpha}=[%
\begin{array}
[c]{ccc}%
\alpha_{x} & \alpha_{y} & \alpha_{z}%
\end{array}
]$. Both $\boldsymbol{C}$ and $\boldsymbol{B}$ depend on the lattice symmetry
\cite{birss,mason51,mason54}.

Minimizing the free energy as $\partial F/\partial\boldsymbol{e}=0$ and
considering that $c_{ij}=c_{ji}$ for the components of the elastic tensor, we
obtain
\begin{align}
\boldsymbol{e}  &  =-\boldsymbol{C}^{-1}\cdot\boldsymbol{B},\label{d-tns}\\
F_{\min}  &  =\boldsymbol{e}^{t}\cdot\boldsymbol{B}/2. \label{f-min}%
\end{align}
The derivation of explicit expressions is straightforward but tedious.
Therefore, we present only those necessary in this work. The stable free
energies for the hexagonal symmetry with $\alpha_{z}=1$ and $\alpha_{x}=1$ are
given as%

\begin{align}
2F_{\min}(\alpha_{z}  &  =1)=\frac{-2c_{33}B_{2}^{2}+4c_{13}B_{2}B_{3}%
-(c_{11}+c_{12})B_{3}^{2}}{(c_{11}+c_{12})c_{33}-2c_{13}^{2}},\label{Fz}\\
2F_{\min}(\alpha_{x}  &  =1)=\frac{-(c_{11}c_{33}-c_{13}^{2})B_{1}^{2}%
}{(c_{11}-c_{12})\{(c_{11}+c_{12})c_{33}-2c_{13}^{2}\}}, \label{Fx}%
\end{align}
respectively. The MAE caused by the ME effects in uniaxial ferromagnets is
given by
\begin{equation}
\Delta K_{u}\equiv F_{\min}(\alpha_{x}=1)-F_{\min}(\alpha_{z}=1).
\end{equation}

The expressions for MS are obtained by inserting Eq. (\ref{d-tns}) into
\begin{align}
\lambda &  \equiv\delta\ell/\ell=e_{xx}\beta_{x}^{2}+e_{yy}\beta_{y}%
^{2}+e_{zz}\beta_{z}^{2}\nonumber\\
&  +e_{xy}\beta_{x}\beta_{y}+e_{yz}\beta_{y}\beta_{z}+e_{zx}\beta_{z}\beta
_{x}, \label{ms}%
\end{align}
where $\delta\ell$ is the variation in the lattice along the direction of the
deformation vector $\boldsymbol{\beta}=[\beta_{x}\ \beta_{y}\ \beta_{z}]$. The
resultant expression of $\lambda$ agrees with that obtained by Clark
\cite{clark80} (see Appendix B). Although $\lambda$ generally depends on both
$\boldsymbol{\alpha}$ and $\boldsymbol{\beta}$, we consider two cases in which
the magnetization is aligned with the $c-$axis: under a lattice deformation
along the $c-$axis ($\boldsymbol{\beta}=[001]$) or within the $a-b$ plane
($\boldsymbol{\beta}=[\beta_{x}\ \beta_{y}\ 0]$). These are denoted as
$\lambda_{c}$ and $\lambda_{a}$, respectively. The resultant expressions are
given as
\begin{align}
\lambda_{c}  &  =\frac{2c_{13}B_{2}-(c_{11}+c_{12})B_{3}}{(c_{11}%
+c_{12})c_{33}-2c_{13}^{2}},\label{Lamda-c}\\
\lambda_{a}  &  =\frac{-c_{33}B_{2}+c_{13}B_{3}}{(c_{11}+c_{12})c_{33}%
-2c_{13}^{2}}. \label{Lamda-a}%
\end{align}

The experimental values of MS for YCo$_{5}$ are presented in TABLE I. Because
$\lambda_{a}=0$, we obtain
\begin{align}
B_{2}  &  =(c_{13}/c_{33})B_{3},\label{B2}\\
\lambda_{c}  &  =-B_{3}/c_{33}, \label{Lc}%
\end{align}
from Eqs. (\ref{Lamda-c}) and (\ref{Lamda-a}). It is noteworthy that the
anisotropic MS constant $\lambda_{c}$ is expressed in terms of only $B_{3}$
and $c_{33}$.

Similarly, from Eqs. (\ref{Fz}) and (\ref{B2}), we obtain
\begin{equation}
2F_{\min}(\alpha_{z}=1)=-B_{3}^{2}/c_{33}.
\end{equation}
The expression of $2F_{\min}(\alpha_{x}=1)$ cannot be simplified. However, by
omitting $c_{12}$, we obtain
\begin{equation}
2F_{\min}(\alpha_{x}=1)\simeq-\frac{B_{1}^{2}}{c_{11}}\frac{c_{11}%
c_{33}-c_{13}^{2}}{c_{11}c_{33}-2c_{13}^{2}}.
\end{equation}
When we also omit the $c_{13}$ terms, the MAE $\Delta K_{u}$ caused by the ME
effects is given as
\begin{equation}
\Delta K_{u}\simeq\frac{1}{2}\left[  -\frac{B_{1}^{2}}{c_{11}}+\frac{B_{3}%
^{2}}{c_{33}}\right]  . \label{DeltaKu}%
\end{equation}
The above expression for $\Delta K_{u}$ appears to be reasonable compared with
the well-known result for the cubic symmetry \cite{chikazumi64},%
\[
\Delta K_{1}=\frac{B_{1}^{2}}{c_{11}-c_{12}}-\frac{B_{2}^{2}}{2c_{44}}.
\]
More accurate estimation could be given by adopting the experimental relation
$c_{12}\sim0.6c_{11}$ for bcc Fe presented in TABLE II. We obtain that
$2F_{\min}(\alpha_{x}=1)\sim-B_{1}^{2}/0.64c_{11}$ in so far as\textbf{
}$c_{11}\gg c_{13}$, because $c_{33}\gtrsim c_{13}$ is expected. In Section
IV, we discuss the quantitative values of $B_{3}$ and $c_{33}$ .

The strain-induced MAE is obtained from the expression of the ME free energy
(Eq. (\ref{Fme})):
\begin{align}
F_{me}(\boldsymbol{\alpha})  &  =B_{1}\left(  \alpha_{x}^{2}e_{xx}+\alpha
_{y}^{2}e_{yy}+2\alpha_{x}\alpha_{y}e_{xy}\right) \nonumber\\
&  +B_{2}\alpha_{z}^{2}\left(  e_{xx}+e_{yy}\right)  +B_{3}e_{zz}\alpha
_{z}^{2}\nonumber\\
&  +2B_{4}\left(  \alpha_{y}\alpha_{z}e_{yz}+\alpha_{x}\alpha_{z}%
e_{xz}\right)  .
\end{align}
We consider two types of lattice deformation: a uniaxial deformation along the
$c-$axis and an in-plane deformation in the $a-b$ plane. For the first
(second) case, we may put $e_{zz}\equiv\delta_{c}$ ($e_{xx}=e_{yy}\equiv
\delta_{a}$). The difference in $F_{me}(\boldsymbol{\alpha})$ for
$\boldsymbol{\alpha}=[1\ 0\ 0]$ and $\boldsymbol{\alpha}=[0\ 0\ 1]$ yields the
variation in the MAE when subjected to small lattice deformation. The MAEs for
the lattice deformation along the $c-$axis and in the $a-b$ plane are given
as
\begin{align}
F_{me}^{c}(100)-F_{me}^{c}(001)  &  \equiv\Delta K_{u}^{c}=-B_{3}\delta
_{c},\label{DKu-c}\\
F_{me}^{a}(100)-F_{me}^{a}(001)  &  \equiv\Delta K_{u}^{a}=\left(
B_{1}/2-B_{2}\right)  \delta_{a},
\end{align}
respectively. The calculation of the MAE under a small lattice deformation is
effective for estimating the contribution to $B_{i}$ from the electronic
structure calculations.

\section{III. Calculation of density of states and magnetic anisotropy energy}

\subsection{A. Methods of calculations}

We adopt a real-space tight-binding (TB) method to calculate the DOS and MAE
for a finite size cluster with 7260 atoms. The Hamiltonian is given by a
$d-$orbital TB model with SOC expressed in terms of the $LS$ coupling. Matrix
elements of the Hamiltonian are given in the Slater--Koster form
\cite{slater54}, and hopping integrals are determined by using the Solid State
Table given by Harrison\textbf{ }\cite{harrison80}. The lattice constants are
considered to be $a=5.0\mathrm{\mathring{A}}$ $\equiv a_{0}$\ and
$c=4.08\mathrm{\mathring{A}\equiv}c_{0}$. These make $c/a=0.816$, which is the
ideal value for the hexagonal lattice. These values are close to the observed
ones, that is, $a=4.94$ and $c=3.98\mathrm{\mathring{A}}$
\cite{andreev95,koudela08}. Near neighbor sites of Co(2c) are taken to be
three Y, three Co(2c), and six Co(3g) sites, and those of Co(3g) sites are
four Y, four Co(2c), and four Co(3g) sites. Among them the shortest distance
is the Co(3g) - Co(3g) pair on the $a-b$ plane. The exchange coupling between
up and down spin states is taken to be 0.064 Ry.\textbf{ }The magnitude of the
$LS$ coupling is considered to be 3.7 and 5 mRy for Co and Y, respectively
\cite{inoue21,mabbs08}.

The MAE is evaluated from the thermodynamic potential $\Omega=E-N\mu$, where
$E,N$, and $\mu$ are the total energy, total electron number, and chemical
potential for d-electrons, respectively. The $d-$electron numbers of Y and Co
are assumed to be 3 and 8.3 per atom, respectively, neglecting $s-$ and
$p-$electrons. The total electron number is 44.5 per f.u. By using Green's
functions $G(\epsilon)$, $\Omega$ is given as,
\[
\Omega=-\frac{1}{\pi}\int_{-\infty}^{\varepsilon_{F}}\operatorname{Im}%
\mathrm{Tr}\ln\boldsymbol{G}(\epsilon)d\epsilon.
\]
$G(\epsilon)$ is expanded up to second order of $LS$ coupling in the
Hamiltonian, and the uniaxial MAE is evaluated by maintaining the
corresponding terms in the expression. The unperturbed Green's functions,
which contain information of lattice structure, are evaluated numerically by
using the orbital symmetrized recursive method expressed in $5\times5$
matrices. \cite{inoue15a,inoue21,inoue87}

\subsection{B. Results for density of states}

\begin{figure}[tb]
\begin{center}
\includegraphics[width=1.0\linewidth]{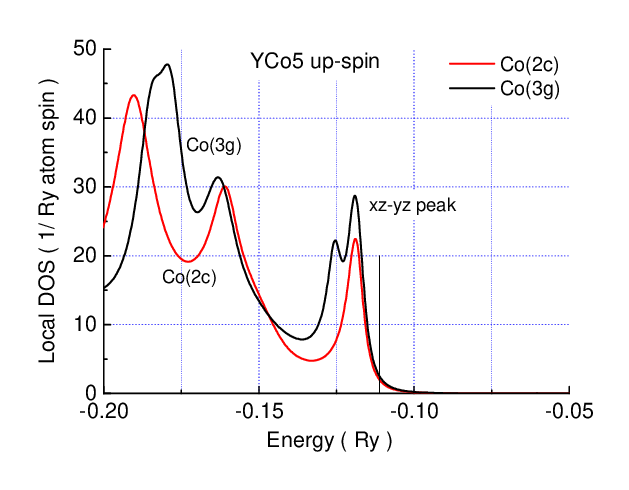}
\end{center}
\caption{(color online) Calculated results of local up-spin density of states
(DOS) of YCo$_{5}$ near Fermi energy $E_{F}=-0.112$ Ry shown by a vertical
line. }%
\label{figure2}%
\end{figure}

Fig. \ref{figure2} shows the calculated results for the local up-spin DOS of
Co(2c) and Co(3g) sites near the Fermi energy $E_{F}=-0.112$ Ry. The position
of $E_{F}$ is shown by a vertical line in the figure. The peaks of the local
DOS appear just below $E_{F}$ for both Co(2c) and Co(3g) sites. It has been
indicated that these DOS peaks are responsible for the MMT.
\cite{yamada99,rosner06,koudela08} To clarify the feature of the DOS peaks, we
performed orbital decomposition of the local DOS in our TB calculations.
Furthermore we calculated the electronic structure in first principles to
confirm the TB results. Details of the results are presented in Appendix C.

Orbital decomposition of the local DOS in TB calculations indicates that the
DOS peak of Co(2c) consists of $xz$ and $yz$ orbitals being originated from
the flat band. \cite{rosner06,koudela08,ochi15} The DOS peaks of Co(3g) appear
to be rather complicated: the main peak consists of $xy$ and $yz$ orbitals,
and the subpeak is formed by $e_{g}$ orbitals, that is, a linear combination
of $x^{2}-y^{2}$ and $3z^{2}-r^{2}$ orbitals. Although the DOS peaks of $xy$
and $yz$ orbitals are nearly degenerate, only the $yz$ orbital-peak is caused
by the flat band. The other peaks are irrelevant to the flat energy bands. The
$xz$ orbital does not contribute to the local DOS of Co(3g). We hereafter call
the sharp DOS peak originated from $xz$ and $yz$ orbitals "$xz-yz$ peak"
according to its orbital character.

The results in first principles agree with those calculated\textbf{
}previously. \cite{rosner06,koudela08} The $xz-yz$ peak is composed of $xz$
and $yz$ orbitals on Co(2c) and of the $yz$ orbital on Co(3g). The local DOS
of $xy$ and $3z^{2}-r^{2}$ orbitals in Co(3g) show no peak structure, but form
broad humps as shown in Fig. \ref{appendix} in Appendix C. Nevertheless, we
may identify that the local DOS of $xy$ and $3z^{2}-r^{2}$ orbitals in first
principles correspond to those of $xy$ and $e_{g}$ orbitals in TB results
according to their dependence on lattice deformation shown in Appendix C.

The appearance of additional peaks in Co(3g) DOS in TB\ calculation might be
attributed to the short atomic distance between the nearest-neighbor Co(3g)
sites on the $a-b$ planes. In our TB calculations, we considered only the
nearest-neighbor atomic pairs for electron hopping between Co(3g) sites on the
$a-b$ planes. Long range electron-hopping may be included effectively in first principles.

The calculated values of the local magnetic moments of Co(2c) and Co(3g) sites
are $1.0$ and $1.2$ $\mu_{B}$ per atom, respectively, and the local magnetic
moment of Y is negative: $-0.22$ $\mu_{B}$. The Co moments are smaller than
the observed ones ($1.31$ and $1.44$ $\mu_{B}$) for Co(2c) and Co(3g),
respectively. Quantitative agreement would be obtained by decreasing the
mixing of wave functions between the Co and Y sites. However, the amendment
would not alter the discussion below.

\begin{figure}[tb]
\begin{center}
\includegraphics[width=1.0\linewidth]{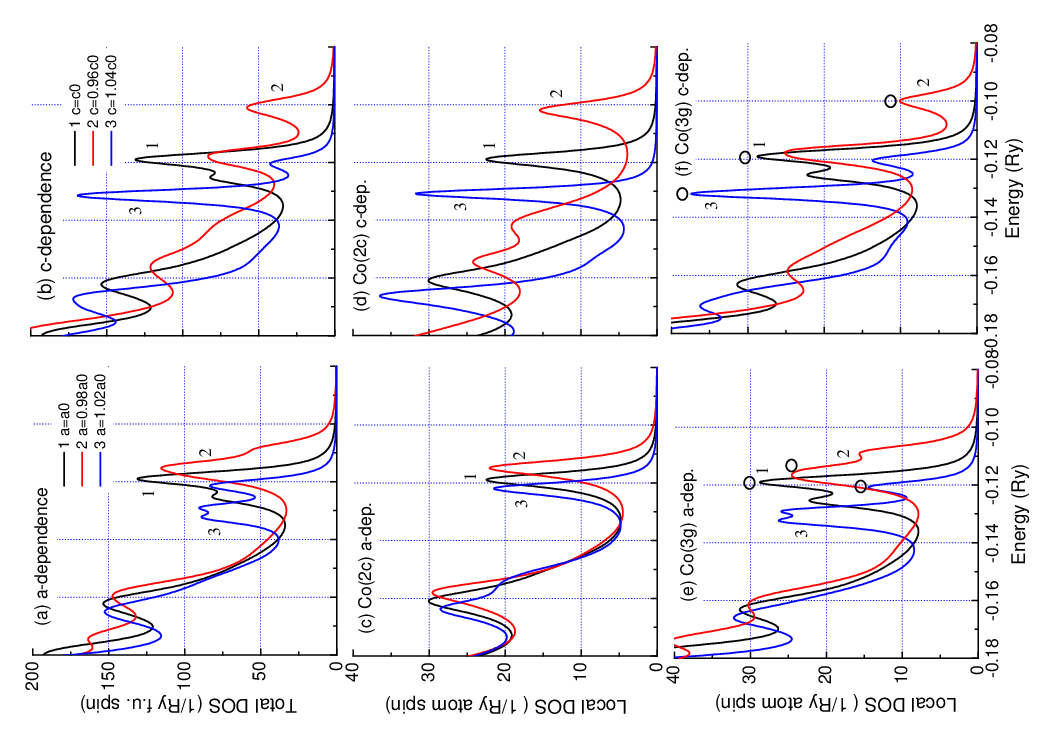}
\end{center}
\caption{(color online) Dependence of up-spin DOS peaks on the variations in
the lattice constants $a$ and $c$, where $a_{0}$ and $c_{0}$ are the lattice
constants without lattice deformation. Results for $a-$dependence of total
DOS, Co(2c) local DOS, and Co(3g) local DOS are shown in (a), (c), and (e),
respectively, and\textbf{ }corresponding results for $c-$dependence are given
in (b), (d), and (f), respectively. The circles in (e) and (f) indicate that
the peaks are related to the $xz-yz$ peak. }%
\label{figure3}%
\end{figure}

Now, we study the deformation dependence of the up-spin DOS peaks by modifying
$a$ and $c$ independently as $a=a_{0}(1\pm\delta)$ or $c=c_{0}(1\pm2\delta)$,
with $\delta=0.01$ and $0.02$. For a fixed value of $\delta$, the lattice
volume is unaltered for $c-$axis and $a-b$ axis deformations in the lowest
order of $\delta$. Figs. \ref{figure3}(a) and (b) show the calculated results
of the dependence of the \textbf{total} up-spin DOS peaks on $a$ and $c$ for
$\delta=$ $0.02$, respectively. The results are decomposed into local ones
calculated for Co(2c) and Co(3g) sites. Figs. \ref{figure3}(c) and (d) show
$a-$ and $c-$dependences of local DOS of Co(2c), respectively, and Figs.
\ref{figure3}(e) and (f) are corresponding results for Co(3g). The results
show that the lattice deformation would not produce a simple shift of DOS
peaks. Rather, it would produce an anisotropic variation in the shape of the
DOS peaks\textbf{ }owing to the reasons explained below.

We start with the results for the local DOS of Co(2c) shown in Figs.
\ref{figure3}(c) and (d). Because the $xz-yz$ peak is the only DOS peak on
Co(2c) in the energy region, the results show simple dependence of the $xz-yz$
peak on the lattice change; it shifts to higher (lower) energy region for
lattice shrinkage (expansion). Because the $xz$ and $yz$ orbitals are
connected strongly along the $c-$axis, the $xz-yz$ peak shifts more
significantly for a $c-$axis change than for an $a-$axis change.

On the contrary, the local DOS peaks of Co(3g) show complicated dependence on
the lattice change as shown in Figs. \ref{figure3}(e) and (f). The $yz$ peak
shows the same dependence with that of the $xz-yz$ peak. The peak position is
identified by the circles added in the figures. The local DOS peaks of $xy$
and $e_{g}$ orbitals shift differently from the $xz-yz$ peak. Because these
orbitals extend more within the $a-b$ planes than along the $c-$axis, they
depend more strongly on $a-$axis change than on $c-$axis change. The tendency
can be seen in Figs. \ref{figure3}(e) and (f).

A superposition of Figs. \ref{figure3}(c) and (e) (\ref{figure3}(d) and (f))
gives the anisotropic dependence of the total DOS on the lattice distortion as
shown in \ref{figure3}(a) (\ref{figure3}(b)). The existence of $xy$ and
$e_{g}$ peaks in the local DOS of Co(3g) may influence the change in magnetic
moments in the MMT as explained below. It also affects the dependence of MAE
on $E_{F}$, which will be discussed in the next subsection.

Under a uniaxial pressure along the $c-$axis, the up-spin DOS may vary as
curve 2 in Fig. \ref{figure3}(b), and $E_{F}$ may be located between two DOS
peaks near $E=-0.12$ and $-0.10$ Ry. The situation has been verified by
self-consistent Hartree--Fock (HF) calculations for YCo$_{5}$ (not shown). The
area of the DOS peak around $E=-0.1$, that is, the electron number occupied in
these states, is approximately 0.57 per f.u. When we assume that the charge
density on the Co sites is unaltered in the MMT, the magnetization variation
$\delta M$ at the MMT is approximately $1.1$ $\mu_{B}$ per f.u. The value of
$\delta M$ is actually consistent with that calculated by Koudela et al.
\cite{koudela08} With a further increase in the pressure, the other peak may
contribute to the MMT, and the total magnetization variation would be
approximately $2\sim2.5\mu_{B}$ per f.u. Thus the present TB calculations
predict a larger value of $\delta M$.

\subsection{C. Results for magnetic anisotropy energy}

\begin{figure}[tb]
\begin{center}
\includegraphics[width=1.0\linewidth]{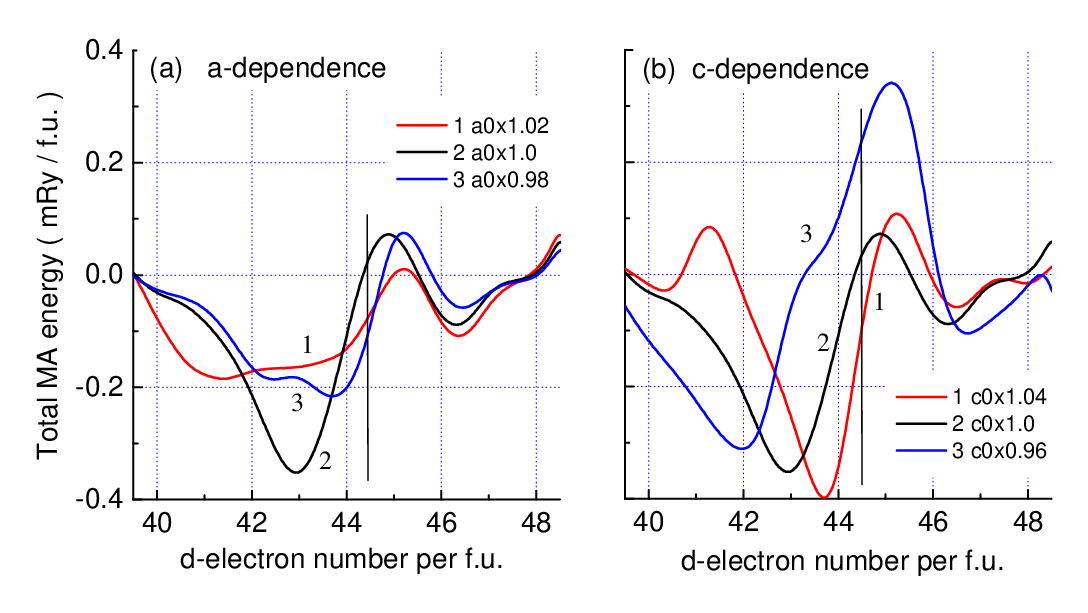}
\end{center}
\caption{(color online) Dependence of MAE on lattice constants (a) $a$ and (b)
$c$, calculated as functions of $d-$electron number $N_{\mathrm{d}}$. $E_{F}$
is located at $N_{\mathrm{d}}=44.5$, as shown by a vertical line.}%
\label{figure4}%
\end{figure}

Figs. \ref{figure4} (a) and (b) show the results of the MAE plotted as a
function of the $d-$electron number per f.u. for the lattice deformations in
the $a-b$ plane and along the $c$-axis, respectively. The total $d-$electron
number assumed in the calculation is 44.5 per f.u. The occupation dependence
of MAE shows an oscillatory dependence as reported previously
\cite{inoue15a,inoue21}. Without lattice deformation, the MAE at
$E_{F}=-0.112$ Ry is $0.034$ mRy/f.u. $=$ 0.47 meV/f.u., which is
significantly smaller than the experimental one. Meanwhile, the large magnetic
anisotropy energy (MAE) in YCo$_{5}$ has been successfully explained in first
principles by incorporating the orbital polarization to reproduce the Hund's
rule coupling. \cite{daalderop96,yamaguchi97,steinbeck01}.

We observe that the MAE depends weakly on the variation in the lattice
constant $a$. However, the detailed dependence of the MAE at the Fermi energy
on the variation in the $a$ values is unclear. This is likely due to the
presence of the up-spin DOS peaks near the Fermi energy and the spin-mixing
term of the SOC. In contrast, the variation in the total MAE for the $c-$axis
distortion shows an anomalous enhancement at $c=0.96c_{0}$. The interpretation
of the enhancement of MAE for $c=0.96c_{0}$ appears to be difficult. However,
a likely interpretation is as follows: When $E_{F}$ is located between the two
separated peaks, the MAE at $E_{F}$ could be the sum of two oscillatory MAE
values near two DOS peaks, and the MAE at $E_{F}$ may become large in a case
where the sum is constructive. The enhancement of MAE has been verified by
self-consistent HF calculations. Thus, the variation in the MAE caused by the
lattice deformation is also anisotropic. A large enhancement of MAE may be
observed when an intermediate state appears in the MMT under lattice
compression along the $c-$axis.

We have shown in the phenomenological theory that $\lambda_{c}=-B_{3}/c_{33}$
for YCo$_{5}$. Once the value of either $B_{3}$ or $c_{33}$ is determined, the
other value is obtained by using the experimental value of $\lambda_{c}$.
Consequently, we are able to estimate the value of $B_{3}^{2}/c_{33}$ which is
a part of the MAE caused by ME effects. In the analysis, however, we should
note the following points: The deformation dependence of the calculated MAE
could be ambiguous due to its oscillatory behavior and possible violation of
Hund's rule coupling. The existence of local DOS peaks in Co(3g) may produce
complicated oscillation in the $E_{F}$ dependence of MAE. We estimate the
values of $B_{3}$ and $c_{33}$ and discuss them in the next Section.

\section{IV. Discussions}

In the previous section, we derived simple expressions of MS and MAE for
YCo$_{5}$. They both are related to $B_{3}$ and $c_{33}$. Here, we estimate
the reasonable values of $B_{3}$ and $c_{33}$ to clarify the origin of the
large value of $\lambda_{c}$. Because no data of these quantities are
available for YCo$_{5}$ at present, let us start by assuming the value of
$c_{33}$.

TABLE II lists the observed\textbf{ }values of the elastic constants of
several ferromagnets. Referring to the Table, we assume $c_{33}\sim
2\times10^{11}$ N/m$^{2}$. Because the observed value $\lambda_{c}%
=6.8\times10^{-3}$ as listed in TABLE I, we obtain $B_{3}=\lambda_{c}\times
c_{33}\simeq-1.4\times10^{9}$ N/m$^{2}$ for YCo$_{5}$. The magnitude of
$B_{3}$ is much larger than that of $B_{1}=1.4\times10^{8}$ N/m$^{2}$ and
$-0.78\times10^{8}$ N/m$^{2}$ estimated for cubic (c-) CoFe$_{2}$O$_{4}$
\cite{suzuki99,lisfi07,inoue14(a)} and tetragonal FePt \cite{inoue13,sakuma94}%
, respectively.

The MAE induced by ME effects is estimated by using Eq. (\ref{DeltaKu}), which
include two terms,\textbf{ }$-B_{1}^{2}/c_{11}$ and $B_{3}^{2}/c_{33}$. When
we assume $B_{1}\sim1\times10^{8}$ N/m$^{2}$ and $c_{11}\sim c_{33}$, we
obtain $B_{3}^{2}/c_{33}\simeq$ $1\times10^{7}$ J/m$^{3}$ and $B_{1}%
^{2}/c_{11}\sim0.5\times10^{5}$ J/m$^{3}$. We find that the contribution from
ME effect to MAE exceeds the observed value $K_{u}=3.8$ meV/f.u.
$=6.9\times10^{6}$ J/m$^{3}$ (1 meV/f.u.=1.812$\times$10$^{6}$ J/m$^{3}$ for
YCo$_{5}$), which is unrealistic.\textbf{ }

\begin{table}[h]
\caption{Values of elastic constants in 10$^{11}$ N/m$^{2}$ reported for
several ferromagnets \cite{chikazumi64,suzuki99,lisfi07}. YFe$_{12}$ and
Y$_{2}$Fe$_{14}$B are estimated using first principles \cite{inoue19}. *A
value for NiFe$_{2}$O$_{4}$ }%
\label{table2}
\begin{center}%
\begin{tabular}
[c]{cccccc}\hline\hline
material & $c_{11}$ & $c_{12}$ & $c_{44}$ & $c_{11}+c_{12}$ & $c_{33}$\\\hline
bcc Fe & 2.41 & 1.46 & 1.22 &  & \\
fcc Ni & 2.50 & 1.60 & 1.18 &  & \\
c-CoFe$_{2}$O$_{4}$ & 2.7 & 1.1 & 0.59$^{\ast}$ &  & \\
tr-Y$_{2}$Fe$_{14}$B &  &  &  & 3.4 & 2.5\\
tr-YFe$_{12}$ &  &  &  & 2.9 & 2.4\\\hline\hline
\end{tabular}
\end{center}
\end{table}

The unrealistic results given above are caused by a large value of $|B_{3}|$
obtained by assuming a standard value of $c_{33}$. Thus, we instead examine
the value of $B_{3}$ first by using the calculated results of the
strain-induced MAE. A simple relation between the strain-induced MAE and ME
constant has been given by Eq. (\ref{DKu-c}) in the phenomenological theory.
Using the results shown in Fig. \ref{figure4}(b), we obtain $\Delta
K_{u}\simeq-3$ meV/f.u. at $\delta_{c}=-0.03$, resulting in $B_{3}%
=-0.6\times10^{8}\ \mathrm{N/m}^{2}$. From the experimental result
$\lambda_{c}(=-B_{3}/c_{33})=6.8\times10^{-3}$, we obtain $c_{33}%
=0.9\times10^{10}\ \mathrm{N/m}^{2}$, and $B_{3}^{2}/c_{33}=0.4\times
10^{6}\ \mathrm{J/m}^{3}$. The magnitude of $B_{3}$ is reasonable as compared
to $|B_{1}|$ observed for $c-$CoFe$_{2}$O$_{4}$ and the tetragonal FePt. The
contribution of ME effects to MAE is small, but it could be non-negligible.

Although the estimated results seem to be reasonable, we should take notice
that the calculated MAE in the TB model is sensitive to the position of
$E_{F}$, and may be enhanced by the oscillatory feature of the MAE. Therefore,
we justify our result by using the calculated results of the strain dependence
of MAE in first principles. The calculated result of MAE dependence on the
lattice constant in first principles \cite{steinbeck01} gives\textbf{ }%
$B_{3}=-1.7\times10^{8}\ \mathrm{N/m}^{2}$, which results in $c_{33}%
=2.5\times10^{10}\ N/m^{2}$ and $B_{3}^{2}/c_{33}=1.2\times10^{6}\ J/m^{3}$ in
the phenomenological analysis.

The difference between the values of $B_{3}$ obtained in TB and
first-principles is only a factor 3. The difference is apparently small,
because of the possible enhancement of MAE by the oscillatory feature in TB
calculations. Nevertheless, we may conclude that both results give the right
order of magnitude for the ME effects in YCo$_{5}$. A noticeable point is that
the value of $c_{33}$ is small compared with those presented\ in TABLE II. The
small value of $c_{33}$ indicates that an elastic softening may occur along
the c-axis in YCo$_{5}$ as pointed out before \cite{koudela08}.

\begin{figure}[tb]
\begin{center}
\includegraphics[width=1.0\linewidth]{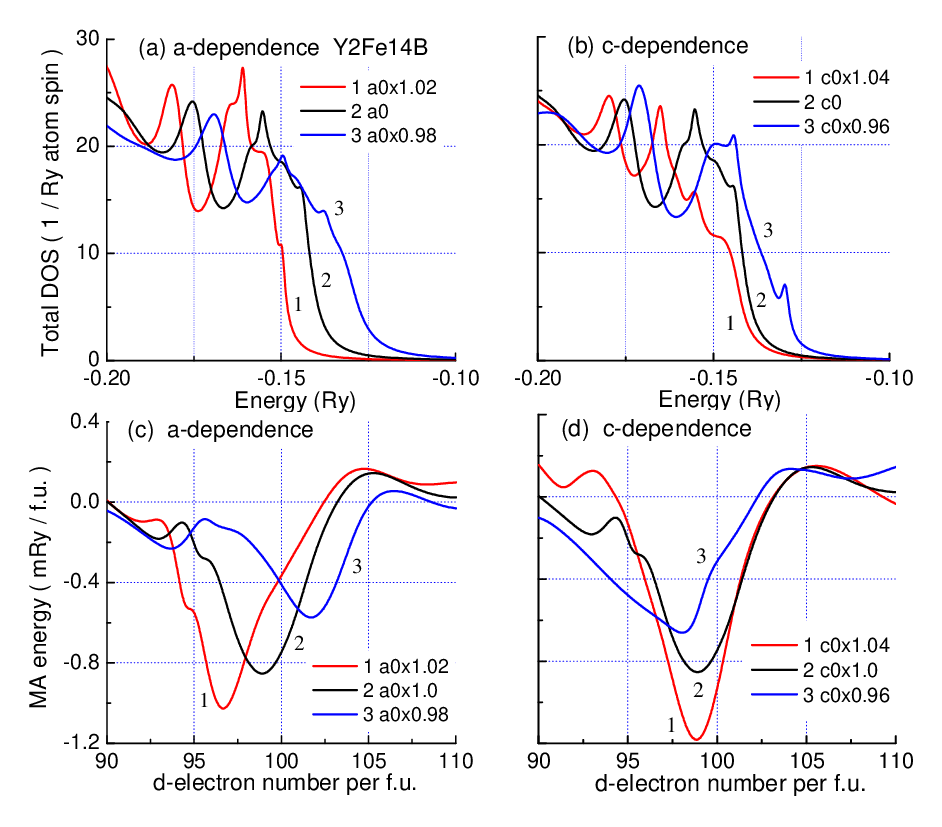}
\end{center}
\caption{(color online) Calculated results of dependence of up-spin DOS of
Y$_{2}$Fe$_{14}$B on lattice constants (a) $a$ and (b) $c$, and corresponding
results of MAE on (c) $a$ and (d) $c$. The number of $d-$electrons in Y$_{2}%
$Fe$_{14}$B is assumed to be 105/f.u. in the calculations. $E_{F}$ is located
at $-0.12$ Ry in undistorted Y$_{2}$Fe$_{14}$B \cite{inoue15b}.}%
\label{figure5}%
\end{figure}

We demonstrated that the anisotropic dependence of the DOS of YCo$_{5}$ on the
lattice distortion is closely related to the feature of the MMT under high
pressure, and that the anisotropic MS is related to an anomaly in the elastic
constant along the $c-$axis. Accordingly, it is anticipated that the
anisotropic MS in the other Y-compounds listed in TABLE I could be interpreted
in a similar manner. Unlike YCo$_{5}$, $\lambda^{c}\ll\lambda^{a}$ appears in
Y$_{2}$Fe$_{14}$B. The relationship is different in the hexagonal Y$_{2}%
$Fe$_{17}$ and rhombohedral Y$_{2}$Fe$_{17}$C$_{1.5}$. To clarify the role of
the lattice structure and electronic states on the anisotropic MS, we
calculated the dependences of DOS and MAE on the lattice deformation for
Y$_{2}$Fe$_{14}$B and h-Y$_{2}$Fe$_{17}$. The calculated results for Y$_{2}%
$Fe$_{14}$B are shown in Fig. \ref{figure5}. Because the $d-$electron number
is considered as 105/f.u. in the calculation, the MS could be affected more by
the variation in $a$ than by that in $c$. This is in accordance with the
experimental observation of MS. However, the results for Y$_{2}$Fe$_{17}$ (not
shown) are complex and do not agree with the experimental results. We
attribute it to the effects of local lattice deformations. \cite{inoue21}

According to the discussions, the softening of $c_{33}$ is crucial for MS in
YCo$_{5}$. So far, the softening of a elastic constant has been proposed to be
a possible origin of the so-called Invar effect \cite{shimizu78,wassermn90}%
.\ As YCo$_{\mathbf{5}}$ also shows the Invar like feature in the temperature
dependence of the lattice constant \cite{andreev95}, YCo$_{5}$ could be a good
candidate to confirm the proposal by calculating $c_{33}$\ in first
principles. Furthermore, it is noticeable that the lattice control of
ferromagnets with large ME effects may enhance their MA. Actually an
extraordinary MA has been observed for epitaxial films of Co ferrite on
Mg$_{2}$SnO$_{4}$ by using the lattice control of thin films
\cite{niizeki13,inoue20b,onoda21}. It is desirable to perform further studies
on ME effects for Y- and RE-intermetallic compounds.

\section{V. Summary}

Using the phenomenological theory, we expressed formulae for the MS and MAE of
uniaxial ferromagnets in terms of the elastic constants and magneto-elastic
constants. The experimental results for MS $\lambda_{a}$ and $\lambda_{c}$ of
YCo$_{5}$ revealed that $\lambda_{c}=-B_{3}/c_{33}$ and that MAE includes a
term of $B_{3}^{2}/c_{33}$ generated by the ME effects. After estimating a
feasible value of $B_{3}$ for YCo$_{5}$, we determined that the elastic
softening predicted earlier should appear in the elastic constant $c_{33}$,
and that the MAE caused by the ME effects may be non-negligible. We showed
that the anisotropic dependence of DOS and MAE on the lattice deformation is
related to the anisotropic MS observed in YCo$_{5}$. It is indicated that the
anisotropic MS observed for various Y-compounds could be interpreted in a
similar manner. We propose further theoretical and experimental studies on ME
effects in related RE intermetallic compounds.

\bigskip

\textbf{Appendix A: Elastic and magnetoelastic constants}

\bigskip

The expressions for the elastic tensors $\boldsymbol{C}$ and vector
presentations of ME constant $\boldsymbol{B}$ and elastic constant
$\boldsymbol{e}$ in Eqs. (\ref{Fel}) and (\ref{Fme}) for the hexagonal
symmetry are given below:%
\[
\boldsymbol{C=}\left[
\begin{array}
[c]{cccccc}%
c_{11} & c_{12} & c_{13} & 0 & 0 & 0\\
& c_{11} & c_{13} & 0 & 0 & 0\\
&  & c_{33} & 0 & 0 & 0\\
&  &  & c_{44} & 0 & 0\\
&  &  &  & c_{44} & 0\\
&  &  &  &  & 2\left(  c_{11}-c_{12}\right)
\end{array}
\right]  ,
\]

\[
\boldsymbol{B}=\left[
\begin{array}
[c]{c}%
B_{1}\alpha_{x}^{2}+B_{2}\alpha_{z}^{2}\\
B_{1}\alpha_{y}^{2}+B_{2}\alpha_{z}^{2}\\
B_{3}\alpha_{z}^{2}\\
2B_{4}\alpha_{y}\alpha_{z}\\
2B_{4}\alpha_{x}\alpha_{z}\\
2B_{1}\alpha_{x}\alpha_{y}%
\end{array}
\right]  ,\boldsymbol{e}=\left[
\begin{array}
[c]{c}%
e_{xx}\\
e_{yy}\\
e_{zz}\\
e_{yz}\\
e_{zx}\\
e_{xy}%
\end{array}
\right]  ,
\]
where $\boldsymbol{\alpha}=[\alpha_{x}\ \alpha_{y}\ \alpha_{z}]$ indicates the
direction cosine of the magnetization vector. For the tetragonal lattice,
$2\left(  c_{11}-c_{12}\right)  $ and $B_{1}\alpha_{x}\alpha_{y}$ should be
replaced with $c_{66}$ and $B_{5}\alpha_{x}\alpha_{y}$, respectively. In the
expression for $F_{me}$, we omit the volume-striction term, $B_{0}\left(
e_{xx}+e_{yy}+e_{zz}\right)  $. This is because it does not include the
direction cosine of the magnetization.

\begin{figure}[tb]
\begin{center}
\includegraphics[width=0.8\linewidth]{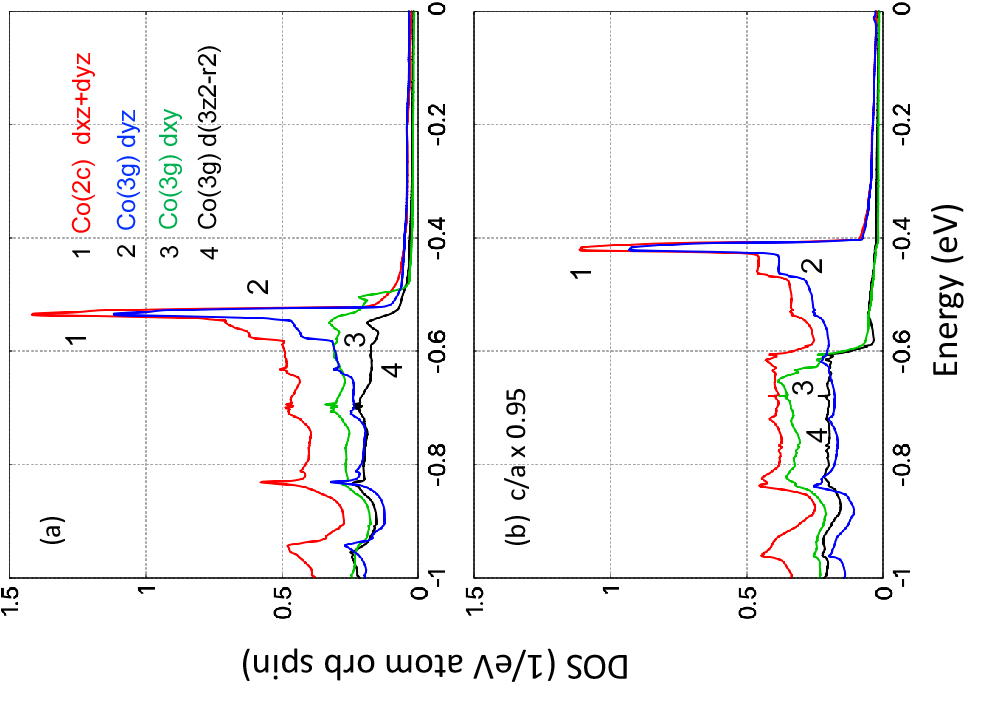}
\end{center}
\caption{(color online) Calculated results of local up-spin DOS of YCo$_{5}$
near Fermi energy $E_{F}$ in first principles. (a) Result for $c/a=0.81$, (b)
those for $c/a=0.77$ with constant lattice volume. }%
\label{appendix}%
\end{figure}

\bigskip

\bigskip

\textbf{Appendix B: Expression of magnetostriction}

\bigskip

The general expression of MS for a hexagonal lattice is derived as%
\begin{align*}
\lambda_{hex}  &  =C_{0}(1-\alpha_{z}^{2})+C_{1}\beta_{z}^{2}+C_{2}(\beta
_{x}^{2}+\beta_{y}^{2})\alpha_{z}^{2}\\
+C_{3}\alpha_{z}^{2}\beta_{z}^{2}  &  +C_{4}\left\{  (\alpha_{x}^{2}%
-\alpha_{y}^{2})(\beta_{x}^{2}-\beta_{y}^{2})+2\alpha_{x}\beta_{x}\alpha
_{y}\beta_{y}\right\} \\
&  +C_{5}(\alpha_{y}\beta_{y}+\alpha_{x}\beta_{x})\alpha_{z}\beta_{z},
\end{align*}
where the coefficients $C_{0}$ $-\ C_{5}$ are given by the lattice constants
$c_{ij}$ and ME constant $B_{i}$. The expression is compared with that
obtained by Clark:
\begin{align*}
\lambda &  =\delta\ell/\ell=\lambda_{1}^{\alpha0}+(\lambda_{2}^{\alpha
0}-\lambda_{1}^{\alpha0})\beta_{z}^{2}+\lambda_{1}^{\alpha2}(\beta_{x}%
^{2}+\beta_{y}^{2})(\alpha_{z}^{2}-1/3)\\
&  +\lambda_{2}^{\alpha2}\beta_{z}^{2}(\alpha_{z}^{2}-1/3)+\lambda^{\gamma
2}\{(1/2)(\beta_{x}^{2}-\beta_{y}^{2})(\alpha_{x}^{2}-\alpha_{y}^{2})\\
&  +2\alpha_{x}\beta_{x}\alpha_{y}\beta_{y}\}+2\lambda^{\varepsilon2}%
(\alpha_{x}\beta_{x}+\alpha_{y}\beta_{y})\alpha_{z}\beta_{z}.
\end{align*}
We observe a good correspondence between these two expressions: $C_{1}%
=\lambda_{2}^{\alpha0}-\lambda_{1}^{\alpha0}$,$\ C_{2}=\lambda_{1}^{\alpha2}%
$,$\ C_{3}=\lambda_{2}^{\alpha2}$, $C_{4}=\lambda^{\gamma2}$, and
$C_{5}=2\lambda^{\varepsilon2},$ except for the term independent of
$\boldsymbol{\beta}$ and omitting $-1/3$. The disagreement regarding the
factor $(1/2)$ in the $C_{4}-$ term may be attributed to a difference in the
definition of the ME coefficients $B_{4}$ and $B_{5}$ by a factor $2$. Note
that these differences are irrelevant in the discussions in the main text.

\bigskip

\begin{figure}[tb]
\begin{center}
\includegraphics[width=0.8\linewidth]{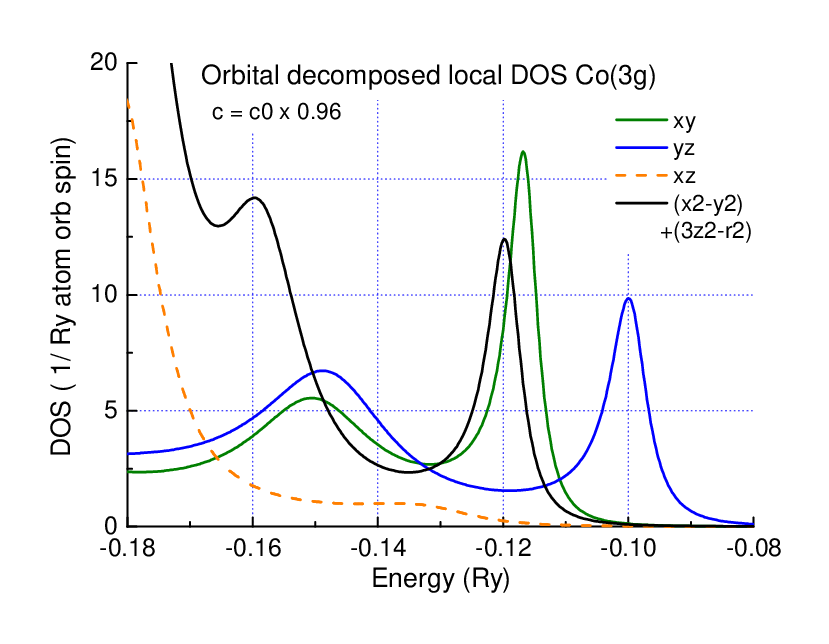}
\end{center}
\caption{(color online) Results of the orbital decomposed local DOS of Co(3g)
for a lattice with $c=c_{0}\times0.96$ in the TB\ calculation. }%
\label{appendix2}%
\end{figure}

\textbf{Appendix C: Calculated results in first principles}

\bigskip

We have performed first principle calculation for YCo$_{5}$ using the
generalized gradient approximation (GGA) method in the WIEN2k package. The
number of $k-$points is $20\times20\times23$ points within the Brillouin zone.
To clarify the effect of lattice deformation on the local DOS, we show the
calculated results for $c/a=0.81$ and $c/a=0.81\times0.95$ in Figs.
\ref{appendix}(a) and (b), respectively. The former results are almost
identical to those calculated previously. \cite{rosner06,koudela08}

As shown in Fig. \ref{appendix}(a), there appear sharp peaks at the top of the
up-spin local DOS of Co, one is assigned to the $xz$ and $yz$ orbitals on
Co(2c) and the other is assigned to the $yz$ orbital on Co(3g). In the local
DOS of Co(2c), no other peak exists, except for these peaks. The contribution
from the $xz$ orbital to the DOS of Co(3g) is quite small (not shown). The
$xy$ and $3z^{2}-r^{2}$ orbitals form no peaks but broad humps near the top of
the up-spin local DOS of Co(3g).

When the $c/a$ is decreased by 5\%, the $xz$ and $yz$ components of the local
DOS shown by numbers 1 and 2 shift to a higher energy region, whereas the $xy$
and $3z^{2}-r^{2}$ components shown by 3 and 4 shift to a lower energy region.
This is because the $c-$axis is compressed while the $a-b$ plane is expanded
in the change of $c/a$ ratio with constant lattice volume.

Fig. \ref{appendix2} shows orbital-decomposed local DOS of Co(3g) in TB
calculation for YCo$_{5}$ with a compressed $c-$axis by 4\%. The corresponding
local DOS of Co(3g) is shown in Fig. \ref{figure3}(f). We find that the $xy$
and $e_{g}$ ($x^{2}-y^{2}$ and $3z^{2}-r^{2}$) components of the local DOS
show peak structures, in addition to the $yz$ component. The contribution from
the $xz$ orbital is negligible for $E>-0.12$ Ry. On the contrary, the $xy$ and
$e_{g}$ components of the local DOS calculated in GGA show no peaks but broad
humps as shown in Fig. \ref{appendix}(b). Nevertheless, the relative position
of the DOS peaks composed of $yz$ and $xy+e_{g}$ orbitals is the same as that
obtained in the GGA calculation. Therefore, we conclude that these DOS peaks
correspond to the local DOS indicated by numbers 3 and 4 in Fig.
\ref{appendix}(b).

The appearance of additional peaks in the local DOS of Co(3g) in the
TB\ calculation might be attributed to the range of electron hopping around
Co(3g) sites specified in the calculation. In the TB calculations, we
considered only the shortest Co(3g)-Co(3g) atomic distance for electron
hopping on the $a-b$ planes. Longer range electron-hopping may be included
effectively in first principles.


\bigskip

\bigskip

\begin{thebibliography}{99}                                                                                               %


\bibitem {kirchmayr79}H. R. Kirchmayr, and C. A. Poldy, \textit{Handbook on
the Physics and Chemistry of Rare Earth}, Vol. 2, edited by K. A. Gshneidner
Jr, and L. Eyring (North-Holland, Amsterdam, 1979), p. 55.

\bibitem {buschow80}K. H. J. Buschow, \textit{Ferromagnetic Materials}, Vol.
1, edited by E. P. Wohlfarth (North-Holland, Amsterdam, 1980), p. 297.

\bibitem {sagawa84}M. Sagawa, S. Fujimura, N. Togawa, H. Yamamoto, and Y.
Matsuura, J. Appl. Phys. \textbf{55}, 2083 (1984).

\bibitem {inoue88}J. Inoue, Physica B \textbf{149}, 376 (1988).

\bibitem {herbst91}J. F. Herbst, Rev. Mod. Phys. \textbf{63}, 819 (1991).

\bibitem {buschow73}K. H. J. Buschow, A. M. van Diepen, and H. W. DeWijn,
Solid State Commun. \textbf{15}, 903 (1973).

\bibitem {sankar75}S. G. Sanker, V. U. S. Rao, E. Segal, W. E. Wallace, W. G.
D.\ Frederick, and H. J. Carret, Phys. Rev. B \textbf{11}, 435 (1975).

\bibitem {alameda83}J. M. Alameda, D. Givord, R. Lemaire, and Q. Lu, J. Magn.
Magn. Mater. \textbf{31}, 191 (1983).

\bibitem {rosner06}H. Rosner, D. Koudela, U. Schwarz, A. Handstein, M.
Hanfland, I. Opahfe, K. Koepernik, M. D. Kuz'min, K.-H. M\"{u}ller, J. A.
Mydosh, and M. Richter, Nat. Phys. \textbf{2}, 469 (2006).

\bibitem {koudela08}D. Koudela, U. Schwartz, H. Rosner, U. Burkhardt, A.
Handstein, M. Hanfland, M. D. Kuz'min, I. Opahle, K. Koepernik, K.-H Muller,
and M. Richter, Phys. Rev. B \textbf{77}, 24411, (2008).

\bibitem {andreev95}A. V. Andreev, \textit{Handbook of Magnetic Materials},
Vol. 8, edited K. H. J. Buschow (Elsevier Science, North Holland, 1995), p. 59.

\bibitem {coehoorn91}R. Coehoorn, J. Magn. Magn. Mater. \textbf{99}, 55 (1991).

\bibitem {nordstrom92}L. Nordstr\"{o}m, M. S. S. Brooks, and B\"{o}rje
Johansson, J. Magn. Magn. Mater. \textbf{104--107}, 1942 (1992).

\bibitem {daalderop96}G. H. O. Daalderop, P. J. Kelly, and M. F. H.
Schuurmans, Phys. Rev. B \textbf{53}, 14415 (1996).

\bibitem {yamaguchi97}M. Yamaguchi, and S. Asano, J. Magn. Magn. Mater.
\textbf{168}, 161 (1997).

\bibitem {steinbeck01}L. Steinbeck, M. Richter, and H. Eschrig, Phys. Rev. B
\textbf{63}, 184431 (2001).

\bibitem {larson03a}P. Larson, and I. I. Mazin, J. Appl. Phys. \textbf{93},
6888 (2003).

\bibitem {larson03b}P. Larson, I. I. Mazin, and D. A. Papaconstantopoulos,
Phys. Rev. B \textbf{67}, 214405 (2003).

\bibitem {yamada99}H. Yamada, K. Terao, F. Ishikawa, Y. Yamaguchi, H.
Mitamura, and T. Goto, J. Phys. Condens. Matter \textbf{11}, 483 (1999).

\bibitem {ochi15}M. Ochi, R. Arita, M. Matsumoto, H. Kino, and T. Miyake,
Phys. Rev. B \textbf{91}, 165137 (2015).

\bibitem {inoue13}J. Inoue, H. Itoh, M. A. Tanaka, K. Mibu, T. Niizeki, H.
Yanagihara, and E. Kita, IEEE Trans. Magn. \textbf{49}, 3269 (2013).

\bibitem {inoue19}J. Inoue, T. Yoshioka, and H. Tsuchiura, IEEE Trans. Magn.
\textbf{55}, 2100304 (2019).

\bibitem {inoue15a}J. Inoue, J. Phys. D: Appl. Phys. \textbf{48}, 445005 (2015).

\bibitem {inoue21}J. Inoue, T. Yoshioka, and H. Tsuchiura, Phys. Rev. Mater.
\textbf{4}, 114404 (2021).

\bibitem {birss}R. Birss, \textit{Symmetry and Magnetism (Selected Topics in
Solid State Physics)}, Vol. 3, edited by E. P. Wohlfarth (North Holland,
Amsterdam, 1964).

\bibitem {mason51}W. P. Mason, Phys. Rev. \textbf{82}, 715 (1951).

\bibitem {mason54}W. P. Mason, Phys. Rev. \textbf{96}, s302 (1954).

\bibitem {clark80}A. E. Clark, \textit{Ferromagnetic Materials}, Vol. 1,
edited by E. P. Wohlfarth (North-Holland, Amsterdam, 1980), p. 531.

\bibitem {chikazumi64}S. Chikazumi, \textit{Physics of Ferromagnetism}, (John
Wiley and Sons, New York, 1964).

\bibitem {slater54}J. C. Slater and G. F. Koster, Phys. Rev. \textbf{94}, 1498 (1954).

\bibitem {harrison80}W. A. Harrison, \textit{Electronic Structure and the
Properties of Solids}, Dover, NewYork (1980).

\bibitem {mabbs08}F. E. Mabbs, and D. J. Machin, \textit{Magnetism and
Transition Metal Complexes}, (Dover, New York, 2008), p. 30.

\bibitem {inoue87}J. Inoue and Y. Ohta, J. Phys. C, Solid State Phys,
\textbf{20}, 1947 (1987).

\bibitem {suzuki99}Y. Suzuki, G. Hu, R. B. van Dover, and R. J. Cava, J. Magn.
Magn. Mater. \textbf{191}, 1 (1999).

\bibitem {lisfi07}A. Lisfi, C. M. Williams, L. T. Nguyen, J. C. Lodder, A.
Coleman, H. Corcoran, A. Johnson, P. Chang, A. Kumar, and W. Morgan, Phys.
Rev. B \textbf{76}, 054405 (2007).

\bibitem {inoue14(a)}J. Inoue, H. Yanagihara, and E. Kita, Mater. Res.
Express, \textbf{1} 046106 (2014).

\bibitem {sakuma94}A. Sakuma, J. Phys. Soc. Jpn. \textbf{63}, 3053 (1994).

\bibitem {inoue15b}J. Inoue, T. Yoshioka, and H. Tsuchiura, J. Appl. Phys.
\textbf{117}, 17C720 (2015).

\bibitem {shimizu78}\textit{The Invar Problem}, Proc. Int. Symp. on the Invar
Problem, Nagoya, Japan, eds. A. J. Freeman and M. Shimizu (North Holland,
Amsterdam, 1978).

\bibitem {wassermn90}E. F. Wasserman, \textit{Ferromagnetic Materials}, Vol.
5, edited by E. P. Wohlfarth and K. H. J. Buschow (North-Holland, Amsterdam,
1990), p. 237.

\bibitem {niizeki13}T. Niizeki, Y. Utsumi, R. Aoyama, H. Yanagihara, J. Inoue,
Y. Yamasaki, H. Nakao, K. Koike, and E. Kita, Appl. Phys. Lett. \textbf{103},
162407 (2013).

\bibitem {inoue20b}J.Inoue, H. Onoda, and H. Yanagihara, JPD, Appl. Phys.
\textbf{53}, 195003 (2020).

\bibitem {onoda21}H. Onoda, H. Sukegawa, J. Inoue, and H. Yanagihara, Adv.
Mater. Interface \textbf{8}, 2101034 (2021).


\end{thebibliography}
\end{document}